\documentclass[twocolumn,pra,superscriptaddress]{revtex4}
\usepackage{amsfonts}
\usepackage{amssymb}
\usepackage{amsmath}
\usepackage{epsfig}
\usepackage{color}
\usepackage{graphics, graphicx}
\usepackage{bbold}
\usepackage{psfrag}
\usepackage{mathcomp}
\usepackage{subfigure}
\usepackage{verbatim}
\usepackage[colorlinks,citecolor=blue]{hyperref}

\makeatletter

\newcommand{\Rmnum}[1]{\expandafter\@slowromancap\romannumeral #1@}
\makeatother

\begin{document}

\title{Topological magnon insulator and quantized pumps \\ from strongly-interacting bosons in optical superlattices}

\author{Feng Mei}
\email{meifeng@sxu.edu.cn}
\affiliation{State Key Laboratory of Quantum Optics and Quantum Optics Devices, Institute
of Laser Spectroscopy, Shanxi University, Taiyuan, Shanxi 030006, China}
\affiliation{Collaborative Innovation Center of Extreme Optics, Shanxi
University,Taiyuan, Shanxi 030006, China}

\author{Gang Chen}
\email{chengang971@163.com}
\affiliation{State Key Laboratory of Quantum Optics and Quantum Optics Devices, Institute
of Laser Spectroscopy, Shanxi University, Taiyuan, Shanxi 030006, China}
\affiliation{Collaborative Innovation Center of Extreme Optics, Shanxi
University,Taiyuan, Shanxi 030006, China}

\author{N. Goldman}
\email{ngoldman@ulb.ac.be}
\affiliation{CENOLI, Universit\'e Libre de Bruxelles, CP 231, Campus Plaine, B-1050 Brussels, Belgium}

\author{Liantuan Xiao}
\affiliation{State Key Laboratory of Quantum Optics and Quantum Optics Devices, Institute
of Laser Spectroscopy, Shanxi University, Taiyuan, Shanxi 030006, China}
\affiliation{Collaborative Innovation Center of Extreme Optics, Shanxi
University,Taiyuan, Shanxi 030006, China}

\author{Suotang Jia}
\affiliation{State Key Laboratory of Quantum Optics and Quantum Optics Devices, Institute
of Laser Spectroscopy, Shanxi University, Taiyuan, Shanxi 030006, China}
\affiliation{Collaborative Innovation Center of Extreme Optics, Shanxi
University,Taiyuan, Shanxi 030006, China}
\date{\today }

\begin{abstract}
We propose a scheme realizing topological insulators and quantized pumps for magnon excitations, based on strongly-interacting two-component ultracold atoms trapped in optical superlattices. Specifically, we show how to engineer the Su-Schrieffer-Heeger model for magnons using state-independent superlattices, and the Rice-Mele model using state-dependent superlattices. We describe realistic experimental protocols to detect the topological signatures of magnon excitations in these two models. In particular, we show that the non-equilibrium dynamics of a single magnon can be exploited to directly detect topological winding numbers and phase transitions. We also describe how topological (quantized) pumps can be realized with magnons, and study how this phenomenon depends on the initial magnon state preparation. Our study opens a new avenue for exploring magnonic topological phases of matter and their potential applications in the context of topological magnon transport.
\end{abstract}

\maketitle

\section{Introduction}

The emblematic Su-Schrieffer-Heeger (SSH) model describes non-interacting fermions hopping on a one-dimensional lattice with alternating hopping matrix elements~\cite{SSH}. This model constitutes a paradigmatic example of a non-trivial topological insulator in 1D, which results from its underlying chiral symmetry~\cite{TOPOKane,TOPOZhang}. It exhibits a variety of topological features, including topological solitons~\cite{Soliton}, fractional charges~\cite{FC}, quantized Zak phases \cite{Zak},  degenerate zero-energy topological edge states \cite{Edge}, and a characteristic ``chiral" displacement \cite{CD}. Adding staggered on-site energies to this model yields the Rice-Mele (RM) model \cite{RM}, which realizes a quantized (Thouless) pump upon varying its parameters in a cyclic and adiabatic manner~\cite{Thouless,Cooper_review}.  The SSH and RM models have both been experimentally realized using \emph{non-interacting single-component} ultracold bosonic atoms trapped in an optical superlattice~\cite{SSHExp}, where quantized and fractional Zak phases were measured. More recently, cold-atom experiments further demonstrated geometric and topological pumping, both for fermionic~\cite{TPFermion} and bosonic gases~\cite{TPBoson,Spielman_pump}, as well as the existence of a topological Anderson insulating phases in the presence of disorder~\cite{Gadway}. Very recently, the SSH model was implemented for hard-core bosons, using Rydberg atoms~\cite{Browaeys}.

On the theoretical side, it was predicted that the SSH model could spontaneously form in systems of interacting bosons described by an extended Hubbard model~\cite{Cuadra}. Besides, the interacting fermionic SSH model was shown to exhibit unusual correlation dynamics upon a quench~\cite{Barbiero}, which can be attributed to a rich interplay between the presence of topological edge modes and spin-charge separation in one dimension.

In this paper, we describe a scheme that realizes the SSH model for magnon excitations, in a system of \emph{strongly-interacting two-component ultracold atoms} trapped in an optical superlattice. Our approach consists in starting from a spinful SSH model for non-interacting two-component ultracold bosonic atoms trapped in a \emph{state-independent} optical superlattice; we then add (Hubbard-type) interactions between the atoms so as to reach the Mott-insulator regime at unit filling (where each lattice site is occupied by a single atom); by making use of a Schrieffer-Wolf transformation, we demonstrate that this interacting system can be described by the SSH model for the underlying magnons (which are bosonic quasiparticle excitations around the ground state of an effective ferromagnet; see Ref.~\cite{Fukuhara2013b} and Eq.~\eqref{mgSSH} below). In this way, we show that the single-particle topological properties of the underlying SSH model are transferred to the bosonic magnonic excitations. Such a transfer of topological band properties to interacting-particle settings is reminiscent of that discussed in Refs.~\cite{Bello,Salerno}, in the context of topological doublons, and in Refs.~\cite{Camacho,Grusdt} in the context of topological polarons.

In the single-magnon excitation subspace, the energy spectrum of the interacting system has two topological magnonic bands, which are characterized by topological winding numbers. Similarly to the more standard single-particle configuration, the system can be prepared in topologically trivial or nontrivial magnonic insulating phases by tuning the tunneling matrix elements of the underlying superlattice. We show that both the value of the topological winding number and the topological phase transition points can be unambiguously detected through nonequilibrium single-magnon quantum dynamics taking place in the optical superlattice. We note that the single-magnon excitation state and its dynamics can be precisely prepared and measured in experiments, based on single-site and time-resolved optical-lattice technologies~\cite{Fukuhara2013b}. Besides, we propose a method based on an array of optical superlattices and using parallel single-magnon state preparation and detection to efficiently measure the single-magnon quantum dynamics.

Furthermore, we show how to promote this system to a RM model for magnons, using a \emph{state-dependent} optical superlattice. We build on the fact that this additional optical superlattice can be controlled dynamically and show that this time-dependent configuration can lead to quantized (topological) pumping of magnons, i.e.~the band structure exhibits non-zero Chern numbers, when taking the dynamical feature of the model into account~\cite{Thouless}. We describe how the shift of the magnon-density center reflects this non-trivial topological invariant, and how the latter depends on the initially prepared single-magnon Bell states. Finally, we discuss how to prepare such Bell states and how to efficiently detect the quantized topological magnon pumping with current optical-lattice technologies.

\section{Su-Schrieffer-Heeger model for magnons}

We consider interacting two-component ultracold bosonic atoms trapped in a
one-dimensional state-independent optical superlattice. As an exemple, one could consider $^{87}$Rb atoms, in which case the two components could be represented by the hyperfine states $|\downarrow \rangle
=|F=1,m_{F}=-1\rangle $ and $|\uparrow \rangle =|F=2,m_{F}=-2\rangle $. The
optical superlattice is generated by superimposing two standing optical
waves that generate a state-independent lattice potential of the form $%
V(x)=V_{l}^{x}\sin ^{2}(k_{1}x)+V_{s}^{x}\sin ^{2}(2k_{1}x+\varphi )$, where
the state-independent lattice potential depths $V_{l,s}^{x}$ and laser phase $%
\varphi $ can be varied by changing the laser power and the optical path
difference. For sufficiently deep optical lattice potential and low
temperature, this optical superlattice system can be described by the
spinful Su-Schrieffer-Heeger-Bose-Hubbard model
\begin{eqnarray}
H_{1} &=&H_{0}+V,  \notag \\
H_{0} &=&-\sum_{x=1}^{N}\sum_{\sigma=\uparrow,\downarrow}\left( J_{1}\hat{a}_{x,\sigma }^{\dag }\hat{b%
}_{x,\sigma }+J_{2}\hat{b}_{x,\sigma }^{\dag }\hat{a}_{x+1,\sigma }+\text{%
H.c.}\right), \notag \\
V&=&\sum_{x=1}^{N}\sum_{\sigma=\uparrow,\downarrow}\frac{U}{2}\left( \hat{a}_{x,\sigma }^{\dag }%
\hat{a}_{x,\sigma }^{\dag }\hat{a}_{x,\sigma }\hat{a}_{x,\sigma }+\hat{b}%
_{x,\sigma }^{\dag }\hat{b}_{x,\sigma }^{\dag }\hat{b}_{x,\sigma }\hat{b}%
_{x,\sigma }\right)  \notag \\
&+&U\sum_{x=1}^{N}\left( \hat{a}_{x\uparrow }^{\dag }\hat{a}_{x\uparrow }\hat{a}%
_{x\downarrow }^{\dag }\hat{a}_{x\downarrow }+\hat{b}_{x\uparrow }^{\dag }%
\hat{b}_{x\uparrow }\hat{b}_{x\downarrow }^{\dag }\hat{b}_{x\downarrow
}\right),
\end{eqnarray}%
where $\hat{a}_{x,\sigma }^{\dag }$ ($\hat{b}_{x,\sigma }^{\dag }$) is the
state-dependent creation operator associated with the lattice site $a_{x}$ ($%
b_{x}$) in the $x$-th unit cell, $J_{1,2}=J\mp \delta J$ are the alternating
tunneling amplitudes, $U$ is the on-site interaction between atoms and $N$ is the unit cell number.
As in recent experiments~\cite{Fukuhara2013a,Fukuhara2013b,Fukuhara2015,Schweizer2016}, we will assume that inter-species and intra-species collisions are characterized by the same interaction strength.

In the limit of strong on-site interaction $U\gg J_{1,2}$, $H_{0}$
(the hopping term) can be seen as a perturbation with respect to $V$ (the on-site
interaction terms). The ground state subspace spanned by the eigenstates of $%
V$ is labeled as $\mathcal{P}$. For unit filling, the ground state
subspace associated with one unit cell can be written as $\mathcal{P}%
=\{|\uparrow,\uparrow \rangle ,|\uparrow, \downarrow \rangle ,|\downarrow,
\uparrow \rangle ,|\downarrow, \downarrow \rangle \}$; these states are associated with the eigenenergy
$E_{g}=0$. The excitation subspace can be be represented by $\mathcal{Q}%
=\{|\uparrow \uparrow ,0\rangle ,|0,\uparrow \uparrow \rangle ,|\uparrow
\downarrow ,0\rangle ,|0,\uparrow \downarrow \rangle ,|\downarrow \downarrow
,0\rangle ,|0,\downarrow \downarrow \rangle \}$, which corresponds to the eigenenergy $%
E_{e}=U$. The projective operators of the above two subspaces are defined as
$\hat{P}=\sum_{|j\rangle \in \mathcal{P}}|j\rangle \langle j|$ and $\hat{Q}%
=\sum_{|k\rangle \in \mathcal{Q}}|k\rangle \langle k|$. Via the
Schrieffer-Wolf transformation \cite{SWT}, the low energy effective
Hamiltonian up to second order is obtained as
\begin{equation}
H^{eff}_{1}=\hat{P}H_{0}\hat{P}+\hat{P}V\hat{P}+\frac{\hat{P}H_{0}\hat{Q}H_{0}%
\hat{P}}{E_{g}-E_{e}}.
\end{equation}%
It is easy to check that $\hat{P}H_{0}\hat{P}+\hat{P}V\hat{P}=0$. For
the intra-cell coupling, after a straightforward calculation, we find that
\begin{eqnarray}
\frac{\hat{P}H_{0}\hat{Q}H_{0}%
\hat{P}}{E_{g}-E_{e}}=- &&\frac{2J_{1}^{2}}{U}%
(|\uparrow, \downarrow \rangle \langle \uparrow, \downarrow |+|\downarrow,
\uparrow \rangle \langle \downarrow, \uparrow |  \notag \\
&+&|\uparrow, \downarrow \rangle \langle \downarrow, \uparrow |+|\downarrow,
\uparrow \rangle \langle \uparrow, \downarrow |)  \notag \\
&-&\frac{4J_{1}^{2}}{U}(|\uparrow, \uparrow \rangle \langle \uparrow, \uparrow
|+|\downarrow, \downarrow \rangle \langle \downarrow, \downarrow |). \notag
\end{eqnarray}%
Similarly, the effective Hamiltonian with respect to the inter-cell coupling
can be derived. Now we introduce the pauli operators $\hat{S}^{1}=|\uparrow
\rangle \langle \downarrow |+|\downarrow \rangle \langle \uparrow |$, $\hat{S%
}^{2}=-i|\uparrow \rangle \langle \downarrow |+i|\downarrow \rangle \langle
\uparrow |$ and $\hat{S}^{3}=(\hat{n}_{\uparrow }-\hat{n}_{\downarrow })/2$.
In terms of these Pauli operators, the total effective Hamiltonian can be
written as a Heisenberg-type model
\begin{align}
H^{\text{eff}}_{1} &=\sum_{x=1}^{N}\biggl[ -\frac{J_{e1}}{2}(\hat{S}_{a_{x}}^{1}\hat{S}_{b_{x}}^{1}+\hat{S}_{a_{x}}^{2}\hat{S}_{b_{x}}^{2})-J_{z1}\hat{S}_{a_{x}}^{3}\hat{S}_{b_{x}}^{3} \notag  \\
& -\frac{J_{e2}}{2}(\hat{S}_{b_{x}}^{1}\hat{S}_{a_{x+1}}^{1}+\hat{S}_{b_{x}}^{2}\hat{S}_{a_{x+1}}^{2})-J_{z2}\hat{S}_{b_{x}}^{3}\hat{S}_{a_{x+1}}^{3}\biggr] , \label{mgSSH}
\end{align}
where $J_{e1,e2}=2J_{1,2}^{2}/U$ and $J_{z1,z2}=4J_{1,2}^{2}/U$ are the
effective spin-exchange couplings.

\begin{figure*}[t]
\includegraphics[width=12cm,height=4.5cm]{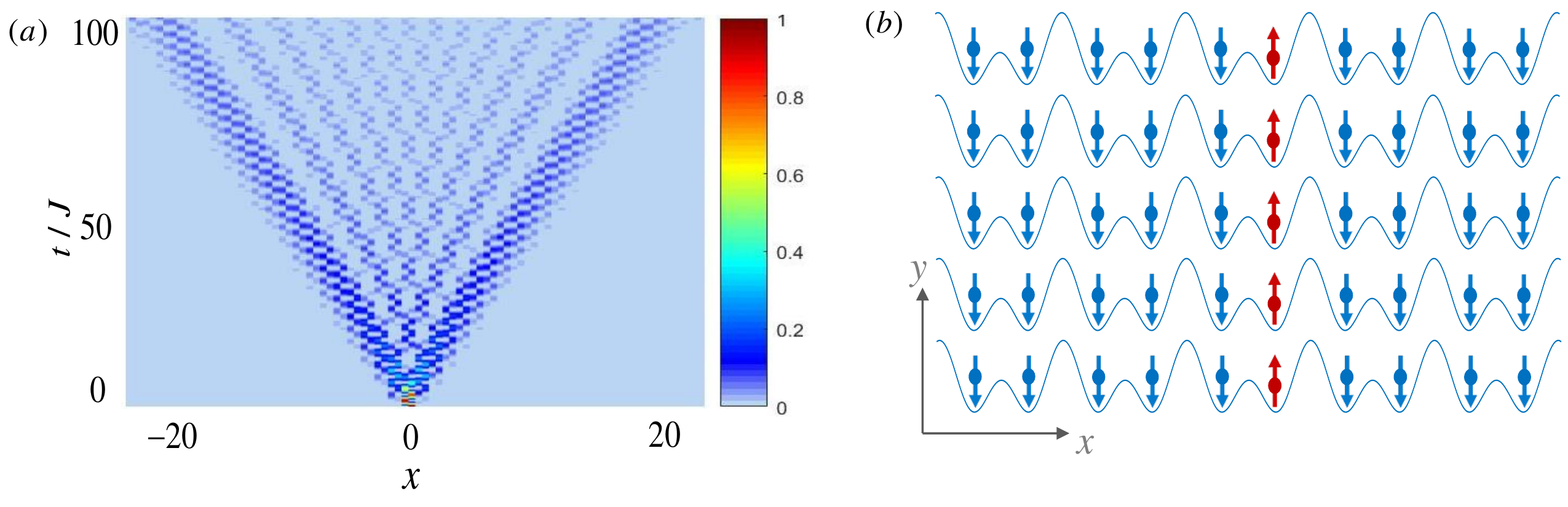}
\caption{(a) Time evolution of the magnon density for a single-magnon state
in an optical superlattice. (b) The schematic diagram of an array of
one-dimensional optical superlattice for highly efficient parallel state
preparation and detection.}
\label{Fig1}
\end{figure*}

In the present work, we restrict ourselves to system configurations that only feature a single spin-up excitation; in practice, this requires temperatures that are low compared to the effective spin-exchange couplings. In this
case, the longitudinal spin coupling in Eq.~(\ref{mgSSH}) only gives rise to
an energy offset and can therefore be neglected. Based on the
Matsubara-Matsuda mapping \cite{MM}, and restricting ourselves to the space of
single spin-up excitations, the above effective spin model can be rewritten in terms of the following magnonic SSH model
\begin{equation}
H_{m1}=\sum_{x=1}^{N}\left( J_{e1}\hat{m}_{a_{x}}^{\dag }\hat{m}%
_{b_{x}}+J_{e2}\hat{m}_{b_{x}}^{\dag }\hat{m}_{a_{x+1}}+\text{H.c.}\right) ,
\label{SSH}
\end{equation}%
where $\hat{m}_{a_{x}(b_{x})}^{\dag }=|\uparrow \rangle
_{a_{x}(b_{x})}\langle \downarrow |$ is the magnon creation operator
associated with the lattice site $a_{x}(b_{x})$ in the $x$-th unit cell, and
$|G\rangle =|\downarrow \downarrow \cdot \cdot \downarrow \downarrow \rangle
$ can be seen as the magnon vacuum state. To study its topological features,
we rewrite it in the momentum-space representation, $\hat{H}_{m1}=\sum_{k_{x}}\hat{m}%
_{k_{x}}^{\dag }\hat{h}(k_{x})\hat{m}_{k_{x}}$, where $\hat{m}_{k_{x}}=(\hat{%
a}_{k_{x}},\hat{b}_{k_{x}})^{T}$, $\hat{a}_{k_{x}}$ and $\hat{b}_{k_{x}}$
are the momentum space operators, $\hat{h}(k_{x})=d_{x}\hat{\sigma}_{x}+d_{y}\hat{\sigma}_{y}$,
where $d_{x}=J_{e1}+J_{e2}\cos (k_{x})$, $d_{y}=J_{e2}\sin (k_{x})$, and $%
\hat{\sigma}_{x}$ and $\hat{\sigma}_{y}$ are the Pauli spin operators
defined in the momentum space. We note that this 2x2 $k$-space Hamiltonian for magnons formally corresponds to that describing the standard SSH model for non-interacting fermions; in particular, it satisfies a chiral symmetry $\Gamma h(k_{x})\Gamma ^{-1}=-h(k_{x})$, where $\Gamma =\hat{\sigma}_{z}$. Besides, the energy spectrum of the system displays two magnonic bands, whose topological features are characterized by the topological winding number
\begin{equation}
\nu =\frac{1}{2\pi }\int dk_{x}\mathbf{n}\times \partial _{k_{x}}\mathbf{n},
\label{wn}
\end{equation}%
where the unit vector is defined as $\mathbf{n}=(n_{x},n_{y})=(d_{x},d_{y})/E$, and where $E=\sqrt{d_{x}^{2}+d_{y}^{2}}$. One reminds that~\cite{Cooper_review}
\begin{equation}
\nu =%
\begin{cases}
1, & \mbox{$J_{1}<J_{2}$}, \\
0\text{,} & \mbox{$J_{1}>J_{2}$}.%
\end{cases}%
\label{tpt}
\end{equation}%
This indicates that the strongly-interacting bosonic system is in the topologically
nontrivial (resp.~trivial) magnonic insulating phase if the hopping matrix elements (or the related spin-exchange
interactions) satisfy $J_{1}<J_{2}$ (resp.~$J_{1}>J_{2}$).

\section{Dynamical detection of topology via single-magnon quantum dynamics}

In contrast to the experimental study performed in the Munich group~\cite%
{Fukuhara2013a}, which investigated quantum dynamics of magnons in an optical lattice, we explore the topological features of
single-magnon quantum dynamics in an optical superlattice. We suppose that the
ultracold atoms that are trapped in the optical superlattice are initially prepared
in the Mott-insulator regime at unit filling:~each optical lattice site traps a single
atom in the internal state $|\downarrow \rangle
=|F=1,m_{F}=-1\rangle $. Based on single-site addressing technology, one can
address a single atom, in the middle of the optical lattice, and flip
its spin into $|\uparrow \rangle =|F=2,m_{F}=-2\rangle $. Then the initial
state of the optical superlattice system is prepared in a single-magnon
state, which can be written as
\begin{equation}
|\psi (0)\rangle =|\downarrow \downarrow \cdots \uparrow \cdots \downarrow
\downarrow \rangle .
\end{equation}%
The quantum dynamics of such single-magnon state is governed by the
single-magnon Hamiltonian in Eq.~(\ref{SSH}). After an evolution time $t$,
the final state of the system becomes
\begin{equation}
|\psi (t)\rangle =e^{-i\hat{H}_{m1}t}|\psi (0)\rangle . \label{smqd}
\end{equation}%
The time evolution of the density distribution of single magnons in the
optical superlattice is shown in Fig.~\ref{Fig1}(a). The results shows that
the single magnon spreads ballistically with two pronounced side peaks and
low probability around the initial position.

Current technologies allow for the time- and site-resolved observation of the
dynamics associated with a single magnon state~\cite{Fukuhara2013a,Fukuhara2013b}. In the following, we will show
the single-magnon quantum dynamics can be efficiently detected based on parallel state preparation
and detection. At the initial time, as shown in Fig. \ref{Fig1}(b), one can prepare a
two-dimensional degenerate Bose gas of $^{87}$Rb atoms in a two-dimensional
optical lattice potential $V(x,y)=V_{l}^{x}\sin ^{2}(k_{1}x)+V_{s}^{x}\sin
^{2}(2k_{1}x+\varphi )+V_{s}^{y}\sin ^{2}(2k_{1}y)$. When $V_{s}^{y}$ is
tuned to be vary large, the hopping along the $y$ direction can be
neglected, and this lattice potential creates an array of independent
one-dimensional optical superlattices along the $x$ direction. In the
Mott-insulator regime, this optical lattice system can be seen as an array
of parallel one-dimensional spin chains. Suppose all the atoms trapped in
the lattice are prepared in the ground state $|\downarrow \rangle
=|F=1,m_{F}=-1\rangle $. Then, an addressing beam profile in form of a line
generated by a spatial light modulator is used to flip the atoms in the
middle lattice sites of all spin chains into the excited state $|\uparrow
\rangle =|F=2,m_{F}=-2\rangle $ \cite{Fukuhara2013a,Fukuhara2013b}. After
that, the single-magnon excitation in each spin chain will propagate with
time. Via rapidly increasing the lattice depth, the single-magnon dynamics
can be frozen to a fixed time and its density distribution can be measured
by single site-resolved fluorescence imaging. The measured magnon density is
obtained by averaging data from the measurements in all spin chains. This
parallel state preparation and detection strategy can greatly improve the
experimental efficiency.

\begin{figure*}[t]
\includegraphics[width=16cm,height=4.5cm]{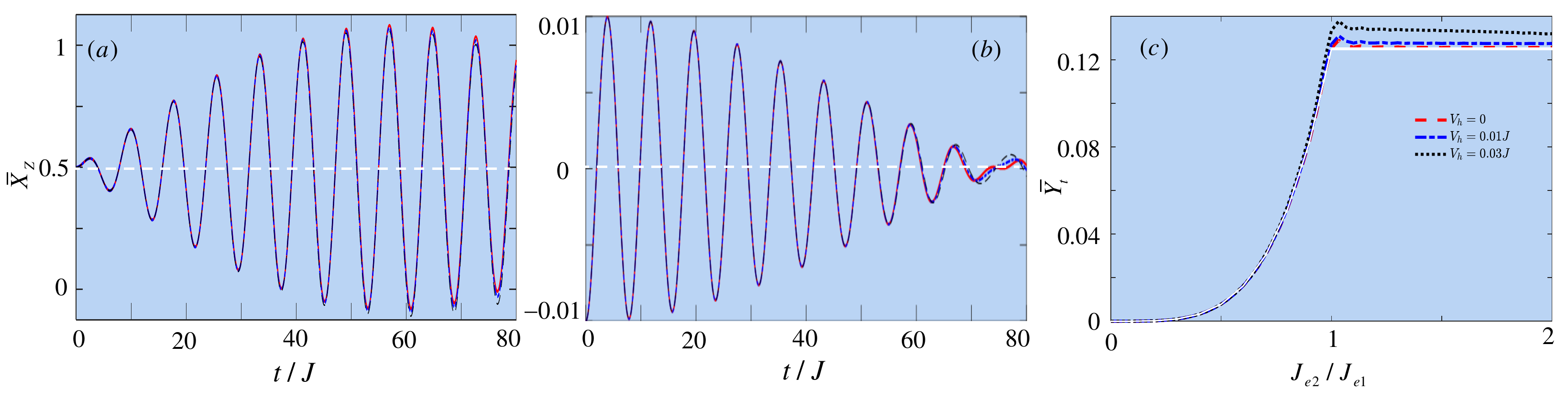}
\caption{Time evolution of the mean chiral displacement for (a)
topological nontrivial insulator phase with $J_{1}=0.2J$ and $J_{2}=J$, and
(b) topological trivial magnon insulator phase with $J_{1}=J$ and $%
J_{2}=0.2J $. (c) The second order moment vs $J_2/J_1$. The evolution time is $t=30/J$. The solid line
curve plots the analytical result shown in Eq. (\protect\ref{gmdc}). The harmonic trap strength is chosen as $V_h=0$ (dashed line), $0.01J$ (dash-dotted line) and $0.03J$ (dotted line). The
other parameters are $J_1=J$, $U=4J$ and $N=43$. $J$ is used as energy unit in this work.}
\label{Fig2}
\end{figure*}

\subsection{Dynamical detection of the topological winding number}

A recent study, performed in the context of linear-optics topological quantum walk, has shown that the topological winding number can be connected with discrete-time quantum dynamics \cite{DTQW2017}. This method has been further generalized to systems with continuous-time dynamics~\cite{CD}. Based on such a method, we show that the topological winding number also can be directly detected through the mean chiral displacement associated with the dynamics of a single-magnon quantum state. For the single-magnon quantum dynamics described in Eq.~(%
\ref{smqd}), the mean chiral displacement is expressed as
\begin{equation}
\bar{X}_{z}(t)=\langle \psi (t)|\hat{X}_{z}|\psi(t)\rangle ,   \label{pdt}
\end{equation}
where the mean chiral displacement operator is defined as $\hat{X}_{z}=\hat{x}\Gamma$. In the long time
limit, a relationship between the winding number and the time-averaged mean chiral displacement can be derived as \cite{DTQW2017}
\begin{equation}
\nu=2\,\mathbb{X}_z,  \label{pdwn}
\end{equation}%
where the time-averaged mean chiral displacement is $\mathbb{X}_z={\lim_{T\rightarrow \infty }}\frac{1}{T}\int_{0}^{T}dt\,\bar{X}_{z}(t)$. As we can see, $\mathbb{X}_z$ is the oscillation center of $\bar{X}_{z}$ varying with time and the topological winding number is twice this center.
Eq.~(\ref{pdwn}) also indicates that the quantum dynamics of a
single-magnon state in a topological optical superlattice can give us the
underling topological information.

To measure the time evolution of the mean chiral displacement based on
single-magnon quantum dynamics in the optical superlattice, we need to
transfer the mean chiral displacement into real space and rewrite it as
\begin{equation}
\hat{X}_{z}=\sum_{x}x\left( \hat{P}_{a_{x}}-\hat{P}_{b_{x}}\right) ,
\end{equation}%
where $\hat{P}_{s}=|\uparrow \rangle _{s}\langle \uparrow |$ ($s=a_{x},b_{x}$%
) is the magnon density operator. By substituting the above equation into
Eq.~(\ref{pdt}), the time evolution of the mean chiral displacement
associated with the dynamics of a single-magnon state described in Eq.~(\ref%
{smqd}) is numerically calculated in Fig.~\ref{Fig2}. The numerical results
in Fig.~\ref{Fig2}(a) show that, when the optical superlattice system is in
the topological nontrivial magnon insulator phase, the mean chiral displacement oscillates around $0.5$. While if the system is in the topological
trivial magnon insulator phase, as shown in Fig.~\ref{Fig2}(b), it
oscillates around $0$. According to Eq.~(\ref{pdwn}), the topological
winding number is twice the oscillation center, which allows one to directly obtain
the topological winding numbers as $\nu =1$ and $0$, respectively.

In addition to the optical superlattice potential, a weak external harmonic
trap is also typically present in experiments. This harmonic trap
introduces a site-dependent potential
\begin{equation}
V_j=V_h(j-j_0)^2,
\end{equation}
where $V_h=\frac{1}{2}m\omega^2$, $m$ is the mass of $^{87}$Rb atoms, $%
\omega $ is the harmonic trap frequency, $j$ is the coordinate of a lattice
site, $j_0 $ is the coordinate of the center of the harmonic trap and the
lattice constant is chosen to be unity. In principle, the super-exchange
couplings will become site-dependent due to the harmonic trap potential.
Specifically, the super-exchange coupling between the $j$-th and $j+1$-th
lattice site can be written as
\begin{equation}
J_{e1,e2}=\frac{2J^2_{1,2}U}{U^2-(V_j-V_{j+1})^2}.
\end{equation}
Besides, the harmonic trap will also cause an on-site energy shift. In Fig.~\ref{Fig2}(a-b), we have numerically analyzed the influence of the external
harmonic trap on the mean chiral displacement by taking into account the
site-dependent super-exchange coupling and the on-site energy shift. Our results show that a weak harmonic trap does not significantly affect the oscillation center of $\bar{X}_{z}$, which indicates that the topological winding number can thus be unambiguously measured in realistic situations.

\subsection{Dynamical detection of the topological-phase-transition point}

In addition to the topological winding number itself, the topological-phase-transition point can also be directly detected via single-magnon quantum dynamics. Indeed, a previous study showed that the topological phase transition of a one-dimensional topological (quantum walk) system can be observed via a second-order moment \cite{DTQW2016}. Building on  this method, here we show that the second order moment associated with the single-magnon continuous-time quantum dynamics, as obtained in the long-time limit, can also be related to the topological phase transitions predicted in Eq.~\eqref{tpt}. The second order moment associated with the dynamics of a
single-magnon state is given by
\begin{equation}
\bar{Y}_{t}=\langle \psi (t)|\hat{Y}_{t}|\psi(t)\rangle,
\label{smdc}
\end{equation}
where the second order moment operator is defined as $\hat{Y}_{t}=\hat{x}^2/t^2$ \cite{DTQW2016}. By transferring the above equation into the momentum space, in the long-time limit, the terms proportional to $1/t$ can be omitted and the above integral can be analytically solved as
\begin{equation}
\bar{Y}_{t}=%
\begin{cases}
\frac{J^2_{e1}}{2}, & \mbox{$J_{1}<J_{2}$} , \\
\frac{J^2_{e2}}{2}, & \mbox{$J_{1}>J_{2}$} . %
\end{cases}
\label{gmdc}
\end{equation}
This equation shows that the topological phase transition point $J_1=J_2$ in
the superlattice Bose-Hubbard model can be precisely measured through the
second order moment in the long-time limit. In this way, the
topological phase transition can also be observed based on single-magnon
quantum dynamics.

The second order moment is extracted from the quantum dynamics
of a single-magnon state in real space. Thus, we rewrite the second order moment operator in the corresponding representation,
\begin{equation}
\hat{Y}_{t}=\sum_{x}x^{2}\left( \hat{P}_{a_{x}}+\hat{P}_{b_{x}}\right)
/t^{2}.
\end{equation}%
After substituting the above equation into Eq.~(\ref{smdc}), we numerically calculate the change of the second-order moment in the long-time limit, and study its behavior as a function of the ratio $J_{2}/J_{1}$; the results are shown in
Fig.~\ref{Fig2} (c). We find that the numerical results for $\bar{Y}_{t}$ agree well with the
analytical result presented in Eq.~(\ref{gmdc}). When $J_{1}$ is fixed and $%
J_{2}>J_{1}$, the second order moment is a constant and will
not change with $J_{2}/J_{1}$, which is reflected by the plateaus in Fig.~%
\ref{Fig2} (c). While if $J_{2}<J_{1}$, the second order moment is a
quadratic function of $J_{2}/J_{1}$. In this way, the topological phase
transition and also the transition point $J_{1}=J_{2}$ can be measured. We
also numerically calculate the influence of the external harmonic trap on
the above topological phase transition signal. These results show that while the
actual value of the plateau is modified by the presence of the harmonic trap,
its position along the $J_2/J_1$ axis remains essentially unchanged. Hence, the topological phase transition can still be unambiguously detected, even in the presence of the trap.

\begin{figure*}[t]
\includegraphics[width=11cm,height=5cm]{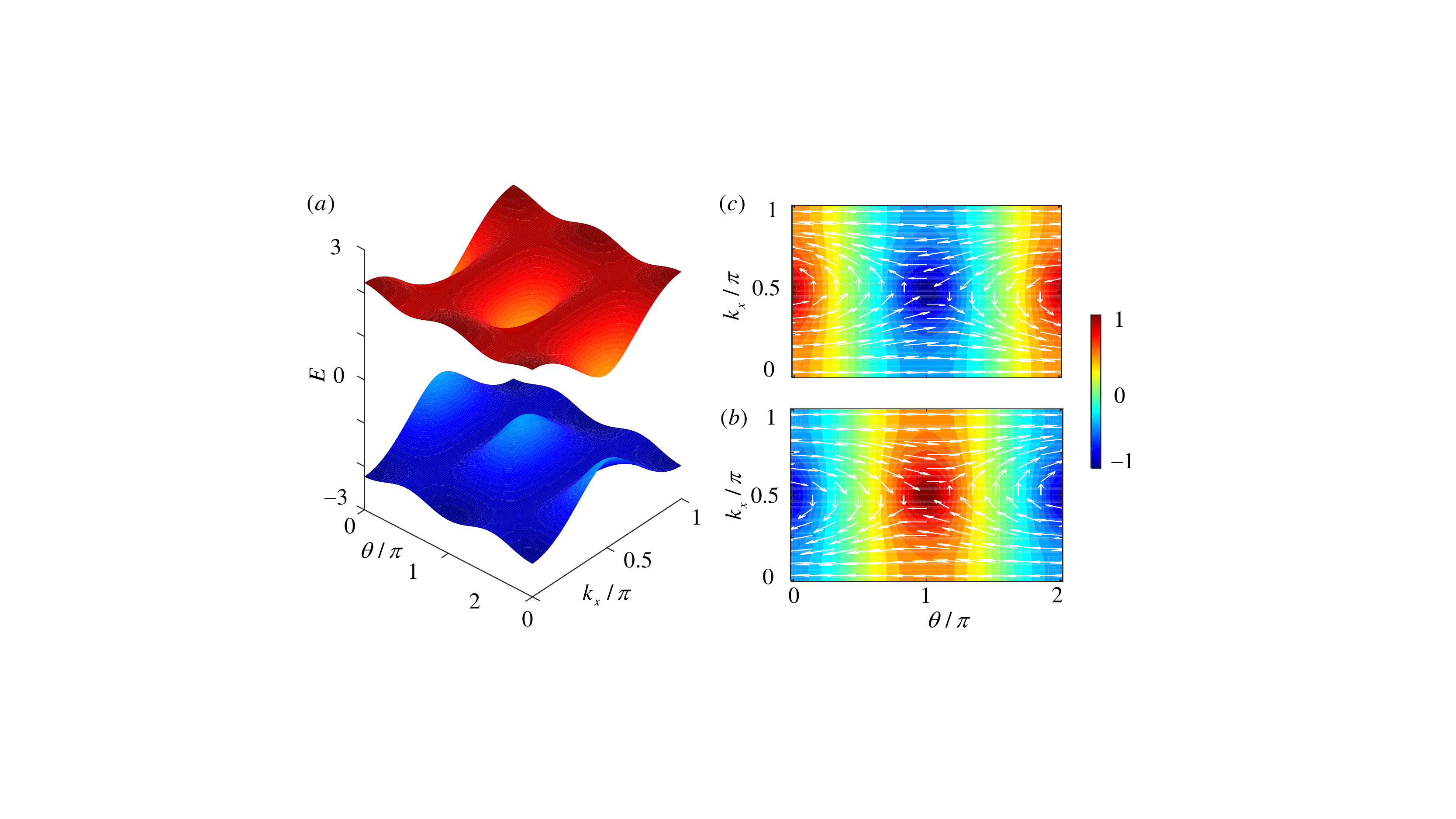}
\caption{(a) The single-magnon energy spectrum with two magnon bands. The
configuration of the unit vector field $\mathbf{n}$ in the first Brillouin
zone for the lower (b) and upper (c) magnon bands. The parameters are chosen
as $J_e=J_p=J$.}
\label{Fig3}
\end{figure*}

\section{Rice-Mele model for magnons}

In this section, we analyze how the RM model and the related topological pump can be obtained by replacing the state-independent optical superlattice by a \emph{state-dependent} one. We point out that such state-dependent optical superlattices have been realized in experiments; see Ref.~\cite{Yuan} for details. The
corresponding potential is taken in the form $V_{\sigma }(x)=V_{l\sigma }\sin
^{2}(k_{1}x)+V_{s\sigma }\sin ^{2}(2k_{1}x+\varphi _{\sigma })$, where the
potential depths $V_{(l,s)\sigma }$ and the phases $\varphi _{\sigma }$
can be varied by changing the laser parameters~\cite{Yuan}.
The corresponding tight-binding Hamiltonian is described by the spinful Rice-Mele-Bose-Hubbard model
\begin{eqnarray}
H_2 &=&H_{0}+V,  \notag \\
H_{0} &=&-\sum_{x=1}^{N}\sum_{\sigma=\uparrow,\downarrow}(J_{1}\hat{a}_{x,\sigma }^{\dag }\hat{b}%
_{x,\sigma }+J_{2}\hat{b}_{x,\sigma }^{\dag }\hat{a}_{x+1,\sigma }+\text{H.c.%
})  \notag \\
&+&\Delta \sum_{x}(\hat{a}_{x\uparrow }^{\dag }\hat{a}_{x\uparrow }-\hat{b}%
_{x\uparrow }^{\dag }\hat{b}_{x\uparrow }-\hat{a}_{x\downarrow }^{\dag }\hat{%
a}_{x\downarrow }+\hat{b}_{x\downarrow }^{\dag }\hat{b}_{x\downarrow }),
\notag \\
V &=&\sum_{x=1}^{N}\sum_{\sigma=\uparrow,\downarrow}\frac{U}{2}(\hat{a}_{x,\sigma }^{\dag }\hat{a}%
_{x,\sigma }^{\dag }\hat{a}_{x,\sigma }\hat{a}_{x,\sigma }+\hat{b}_{x,\sigma
}^{\dag }\hat{b}_{x,\sigma }^{\dag }\hat{b}_{x,\sigma }\hat{b}_{x,\sigma })
\notag \\
&+&U\sum_{x}(\hat{a}_{x\uparrow }^{\dag }\hat{a}_{x\uparrow }\hat{a}%
_{x\downarrow }^{\dag }\hat{a}_{x\downarrow }+\hat{b}_{x\uparrow }^{\dag }%
\hat{b}_{x\uparrow }\hat{b}_{x\downarrow }^{\dag }\hat{b}_{x\downarrow }),
\end{eqnarray}%
where $\Delta $ is the state-dependent staggered on-site
energy. In the strong interaction case, based on the Schrieffer-Wolf transformation \cite{SWT}, the low energy effective Hamiltonian up to second order can be expressed as
\begin{eqnarray}
H^{\text{eff}}_2 &=&\sum_{i}[-\frac{J_{e1}}{2}(\hat{S}_{a_{i}}^{x}\hat{S}_{b_{i}}^{x}+\hat{S}%
_{a_{i}}^{y}\hat{S}_{b_{i}}^{y})-J_{z1}\hat{S}_{a_{i}}^{z}\hat{S}_{b_{i}}^{z}
\notag \\
&-&\frac{J_{e2}}{2}(\hat{S}_{b_{i}}^{x}\hat{S}_{a_{i+1}}^{x}+\hat{S}_{b_{i}}^{y}\hat{S}%
_{a_{i+1}}^{y})-J_{z2}\hat{S}_{b_{i}}^{z}\hat{S}_{a_{i+1}}^{z}]  \notag \\
&+&\sum_{i}\Delta (\hat{S}_{a_{i}}^{z}-\hat{S}_{b_{i}}^{z}),  \label{DSPC}
\end{eqnarray}%
where the alternating spin-exchange couplings can be rewritten as $J_{e1,e2}=(J\mp \delta J)^{2}/U=J_{e}\mp \delta J_{e}$.

In the single-magnon subspace, the above effective spin model can be rewritten into the following magnonic RM model
\begin{eqnarray}
H_{m2} &=&\sum_{i}(J_{e1}\hat{m}_{a_{i}}^{\dag }\hat{m}_{b_{i}}+J_{e2}\hat{m}%
_{b_{i}}^{\dag }\hat{m}_{a_{i+1}}+\text{H.c.})  \notag \\
&+&\sum_{i}\Delta (\hat{m}_{a_{i}}^{\dag }\hat{m}_{a_{i}}-\hat{m}%
_{b_{i}}^{\dag }\hat{m}_{b_{i}}). \label{Hm}
\end{eqnarray}%
Using current ultracold-atom technologies~\cite{TPFermion,TPBoson}, one can modify the optical-lattice potential so as to modulate the parameters $(\delta J_{e},\Delta )$ adiabatically, in view of realizing a closed loop in parameter space~\cite{Thouless,Cooper_review}. In this way, $\big(\delta J_{e},\Delta )$
can be parameterized as $(J_{p}\sin (\theta ),J_{p}\cos (\theta )\big)$,
where $\theta $ defines a dynamical parameter. Then the spin super-exchange
couplings and on-site energy offset can be written as $J_{e1,e2}=J_{e}\mp
J_{p}\sin (\theta )$ and $\Delta =J_{p}\cos (\theta )$. To investigate the
topological features of the magnon Hamiltonian (\ref{Hm}), we write it in
the momentum space as $H_{m2}=\sum_{k_{x}}\hat{m}_{k_{x}}^{\dag }\hat{h}%
(k_{x},\theta )\hat{m}_{k_{x}}$, where $\hat{m}_{k_{x}}=(\hat{m}_{a_{k_{x}}},%
\hat{m}_{b_{k_{x}}})^{T}$. Specifically, the momentum density is written as
\begin{equation}
\hat{h}(k_{x},\theta )=h_{x}\hat{\sigma}_{x}+h_{y}\hat{\sigma}_{y}+h_{z}\hat{%
\sigma}_{z},
\end{equation}%
where $h_{x}=2J_{e}\cos (k_{x})$, $h_{y}=2J_{p}\sin (\theta )\sin (k_{x})$
and $h_{z}=J_{p}\cos (\theta )$. $\hat{\sigma}_{x,y,z}$ are the Pauli
matrixes spanned by $\hat{m}_{a_{k_{x}}}$ and $\hat{m}_{b_{k_{x}}}$.

As for the standard RM model~\cite{Cooper_review}, one can construct a two-dimensional (artificial) Brillouin zone
spanned by the momentum $k_{x}\in (0,\pi ]$ and the dynamical parameter $%
\theta \in (0,2\pi ]$. The single-magnon energy spectrum is plotted in Fig. %
\ref{Fig3}(a), which has two magnonic bands. The topological features of these
two magnon bands are characterized by the Chern numbers. Based on a mapping
from the momentum space to an unit sphere, i.e., $\mathbb{T}^{2}\rightarrow S^{2}$,
the Chern number can be defined as \cite{TOPOKane,TOPOZhang}
\begin{equation}
C=\frac{1}{4\pi }\int \int dk_{x}d\theta (\partial _{k_{x}}\mathbf{n}\times
\partial _{\theta }\mathbf{n})\cdot \mathbf{n},
\end{equation}%
where the unit vector field $\mathbf{n}=(h_{x},h_{y},h_{z})/h$ with $h=\sqrt{%
h_{x}^{2}+h_{y}^{2}+h_{z}^{2}}$. The integrand $\mathbf{n}\times \partial
_{\theta }\mathbf{n}\cdot \mathbf{n}$ is simply the Jacobian of the aforementioned
mapping. Its integration is a topological winding number giving the total
area of the image of the Brillioun zone $\mathbb{T}^{2}$ on $S^{2}$ \cite%
{TOPOKane,TOPOZhang}. It means that, when $(k_{x},\theta )$ wraps around the
entire first Brillouin zone $T^{2}$, this winding number is equal to the
number of times the vector $\mathbf{n}$ wraps around the unit sphere $S^{2}$%
, which is independent of the details of the band structure parameters.

For the sake of completeness, we plot the unit vector
configurations $(n_{x},n_{y})$ and the contours of $n_{z}$, for the lower and
upper magnon bands, in Figs.~\ref{Fig3}(b)-(c), respectively. The results show that, for the lower
(upper) magnon band, the unit vector $\mathbf{n}$ starts from the North
(South) pole at the Brillouin zone center and ends at the South (North) pole
at the Brillouin zone boundary after wrapping around the unit sphere once.
Thus, the Chern number corresponding to the lower (upper) magnon band corresponds to $C_{l}=1$ ($C_{u}=-1$). Since $\theta $ is a periodic dynamical
parameter introduced to construct the first Brillouin zone, the above
nontrivial Chern numbers characterize a dynamical version of a topological
magnon insulating phase.

\begin{figure*}[t]
\includegraphics[width=7.5cm,height=5.5cm]{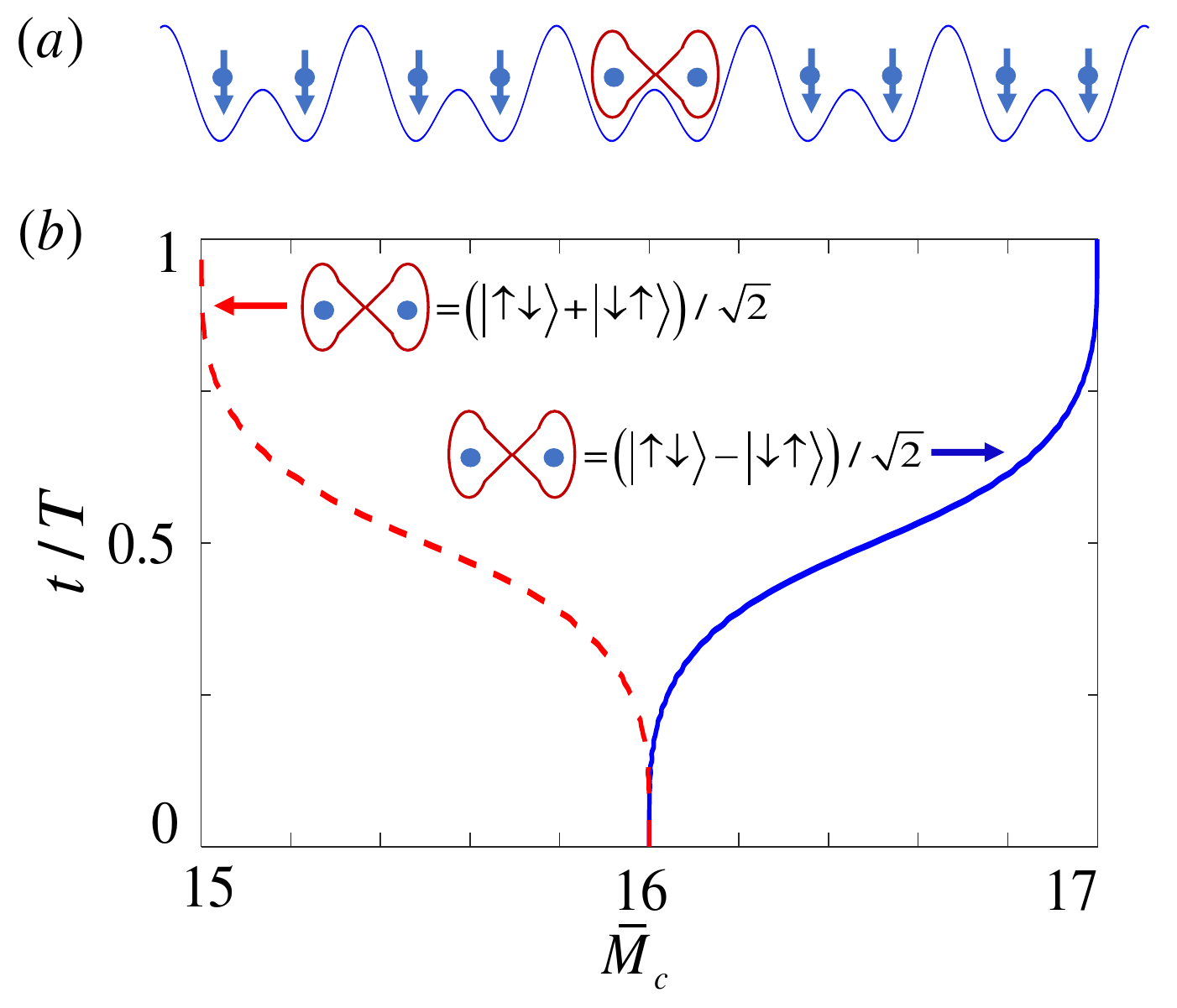}
\caption{(a) The schematic diagram of a one-dimensional optical superlattice
system prepared in a single-magnon state. (b) The change of the magnon
density center $\bar{M}_c$ versus time for different initial spin entangled
state. The other parameters are chosen as $J_e=J_p=J$, $\Omega=0.1J$ and $N=62$.}
\label{Fig4}
\end{figure*}

\section{Quantized topological magnon pumping}

This section describes how the dynamical control over the topological magnonic bands allows for the implementation of a quantized pump for magnons. As described below, this requires a special preparation of the initial single-magnon state. The pump itself is then realized by adiabatically
tuning the dynamical parameter $\theta=\Omega t+\theta_0$ over one period, where $%
\Omega$ is the modulation frequency and $\theta_0$ is the initial phase.

At the initial time, we assume the whole optical superlattice system
consists of series of independent double wells by tuning the optical
superlattice potential to make $J_{e2}=0$. This is equivalent to require the
initial periodic parameter $\theta (t=0)=\theta _{0}=-\text{arcsin}%
(J_{e}/J_{p})$. Suppose the initial system stays in the magnon vacuum state $%
|G\rangle =|\downarrow \downarrow \cdots \downarrow \downarrow \rangle $.
Note that the Hamiltonian for the single-magnon excitation in each double
well has two eigenstates
\begin{equation}
|\chi _{l,u}\rangle =\frac{1}{\sqrt{2}}\left( |\uparrow \downarrow \rangle
\mp |\downarrow \uparrow \rangle \right) ,
\end{equation}%
which are two different Bell states. As shown in Fig. \ref{Fig4}(a), suppose
that one of middle double wells is prepared into the Bell states $|\chi
_{l}\rangle $, while the other qubits stay in the ground state. The initial
state of the system then can be described by $|\psi _{l}\rangle =|\downarrow
\downarrow \cdots \chi _{l}\cdots \downarrow \downarrow \rangle $. Actually,
one can find that $|\psi _{l}\rangle $ ($|\psi _{u}\rangle $) is just the Wannier function
corresponding to the lower (upper) magnon bands. The reason is that the coupling
between nearest neighbor double wells is zero and so the Wannier function
only localizes the middle double well.

After having prepared the initial single-magnon state, the parameter $\theta$
is adiabatically modulated from $\theta(t=0)$ to $\theta(t=T=2\pi/\Omega)$,
the single-magnon wave packet will experience an adiabatic transfer. We
employ a magnon density center to monitor this transfer. Specifically, the
operator for such a center is defined as
\begin{equation}
\hat{M_c}=\sum^{N}_{x=1} x(\hat{P}_{a_x}+\hat{P}_{b_x}),
\end{equation}
where $\hat{P}_{a_x(b_x)}=\hat{m}_{a_x(b_x)}^\dag\hat{m}_{a_x(b_x)}=|%
\uparrow\rangle_{a_x(b_x)}\langle \uparrow|$ is the magnon density in the
lattice site $a_x$ ($b_x$). Then, when the parameter $\theta $ is
adiabatically modulated, the corresponding magnon density center can be
written as
\begin{eqnarray}
\bar{M}_{c}(\theta)&=&\langle \psi _{l}(\theta)|\hat{M_c}|\psi
_{l}(\theta)\rangle  \notag \\
&=&\frac{1}{2\pi }\int dk_{x}i\langle u_{k_{x},\theta,l}|\partial
_{k_{x}}|u_{k_{x},\theta,l}\rangle  \notag \\
&=&\frac{1}{2\pi }\int dk_{x}A_{l}(k_{x},\theta ).
\end{eqnarray}
It turns out that the magnon density center is linked to the Berry
connection $A_{l}(k_{x},\theta )=i\langle u_{k_{x},\theta ,l}|\partial
_{k_{x}}|u_{k_{x},\theta ,l}\rangle $. Therefore, the magnon density center
depends on the gauge choice of the Bloch state. However, the change of the
magnon density center is gauge invariant and thus can be well defined.

Suppose the periodic parameter $\theta $ is changed continuously from $\theta
_{i}$ to $\theta _{f}$. The resulting magnon density center shift is
\begin{equation}
\delta M_c=\bar{M}_{c}\left( \theta _{f}\right) -\bar{M}_{c}\left( \theta _{i}\right) =%
\frac{1}{2\pi }\int dk_{x}\big(A_{l}(k_{x},\theta _{f})-A_{l}(k_{x},\theta
_{i})\big).  \label{MS}
\end{equation}%
By means of the Stokes theorem, the formula (\ref{MS}) can be rewritten as
an integral of the Berry curvature $F_{l}(k_{x},\theta )$ over the surface
spanned by $k_{x}$ and $\theta $, where $F_{l}(k_{x},\theta )=\nabla \times
A_{l}(k_{x},\theta )=i\left( |\langle \partial _{\theta }u_{k_{x},\theta
,l}|\partial _{k_{x}}u_{k_{x},\theta ,l}\rangle -\text{c.c.}\right) $. For a
periodic cycle, $\theta _{f}=\theta _{i}+2\pi $, $H(\theta _{i})=H(\theta
_{f})$, and the change of the magnon center over one cycle is given by the
integral of the Berry curvature over the torus $\{k_{x}\in (0,\pi ],\theta
\in (0,2\pi ]\}$. It is easy to check that the magnon center shift in this
case is just the Chern number of the lower magnon band, i.e.,
\begin{eqnarray}
\delta M_c &=&\frac{1}{2\pi }%
\int_{k_{x}}\int_{\theta }dk_{x}d\theta \nabla \times A_{l}(k_{x},\theta )
\notag \\
&=&\frac{1}{2\pi }\int_{k_{x}}\int_{\theta }dk_{x}d\theta
\,F_{l}(k_{x},\theta )  \notag \\
&=&C_{l}.
\end{eqnarray}%
Similarly, if the initial single-magnon excitation in the middle double well
is prepared in the Bell state $|\chi _{u}\rangle $, after tuning the
parameter $\theta $ over one period, the shift of the magnon density center
becomes $C_{u}$. Therefore, the quantized topological magnon pumping depends on the initial internal spin entanglement $|\chi _{l,u}\rangle $.

The detailed performance of the above topological pumping was numerically
calculated and shown in Fig.~\ref{Fig4}(b). When the two atoms in the middle double
well are prepared in the Bell state $|\chi _{l}\rangle $ ($|\chi
_{u}\rangle $), then the numerical results show that the magnon density center is
shifted to the right (left) by one unit cell after one pumping cycle,
which is indeed compatible with the Chern number of the lower (upper) magnon band $C_{l}=1$ ($%
C_{u}=-1$). Such a quantized pump is robust with respect to the presence of external harmonic trap. With $V_h=0.03J$ and $U=4J$, numerical calculation shows that the resulted fluctuation on $\delta M_c$ caused by the external harmonic trap is $0.08$. Importantly, the numerical results confirm that the Bell state configuration inside
the initial magnon wave packet establishes whether the ground or the excited magnonic bands are occupied, and they illustrate the corresponding band-dependent quantized topological
pumps.

\begin{figure*}[t]
\includegraphics[width=11cm,height=4.5cm]{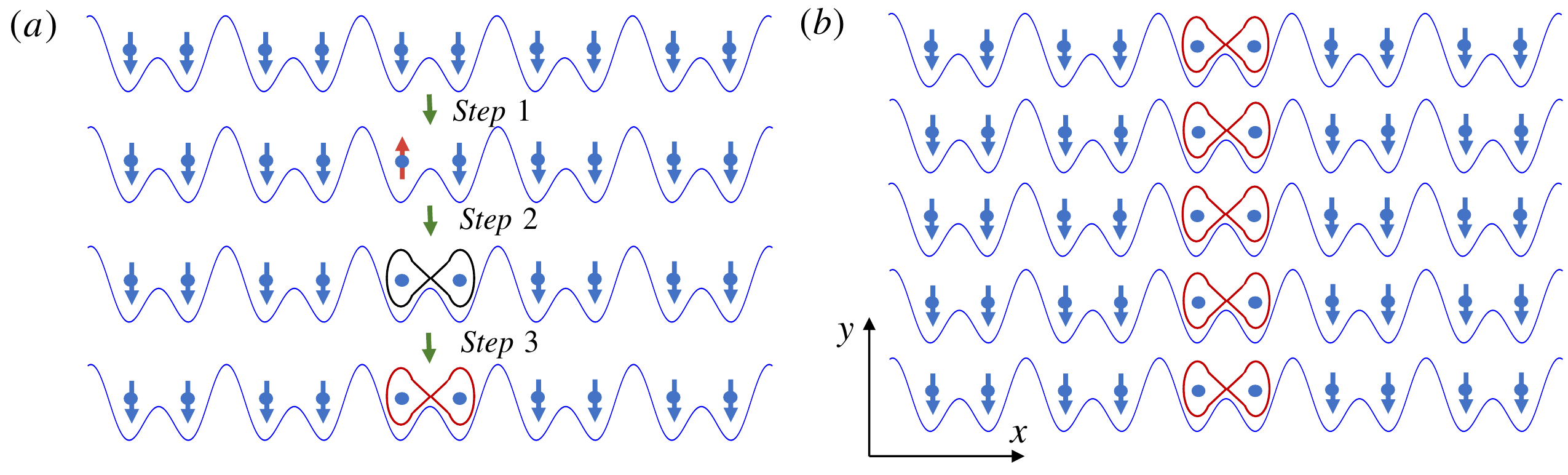}
\caption{(a) The procedure for preparing a spin entangled state in the
middle double well. (b) An array of one-dimensional state-dependent optical
superlattices for highly efficient parallel entangled state generation and
magnon density detection.}
\label{Fig5}
\end{figure*}

\section{Initial state preparation and parallel topological pumping}

The initial single-magnon state with internal spin entanglement can be
prepared based on the spin-exchange Hamiltonian and an effective magnetic
field. This method was recently implemented in a
state-dependent optical superlattice system~\cite{Yuan}. The detailed
preparation procedure is shown in Fig. \ref{Fig5}(a). Initially, the
superlattice potential is tuned so as to impose $J_{e2}=0$ and $\Delta=0$. The
resulting lattice is formed by an array of independent double wells. Suppose
the system is initially prepared in the Mott-insulator regime where each
lattice site has a single atom prepared in the state $|\downarrow \rangle$.
The initial single-magnon state can be prepared via three steps. Step 1:
based on single-site microwave pulse addressing, one of the spins in the
middle double well is flipped into $|\uparrow\rangle$, then the state of the
two spins in the middle double well becomes $|\uparrow\downarrow\rangle$.
Step 2: through a dynamical evolution governed by the spin superexchange
Hamiltonian in Eq.~(\ref{DSPC}) with $\Delta=0$ and evolution time $%
t=\pi/4J_{e1}$, an entangled state $(|\uparrow \downarrow \rangle
-i|\downarrow \uparrow \rangle)/\sqrt{2}$ is generated between the two spins
in the middle double well. Step 3: one creates an effective magnetic field (as described by the last terms of the Hamiltonian in Eq.~\eqref{DSPC}), by switching on the parameter $\Delta$; this step is realized by tuning the state-dependent optical
superlattice potential. This effective magnetic field modulates the phase of the entangled state, which allows one to generate the desired Bell state $%
(|\uparrow \downarrow \rangle \pm|\downarrow \uparrow \rangle)/\sqrt{2}$. During the whole process, one assumes that the state of the spins in the other
double wells remain unchanged. In this way, the single-magnon state $|\psi
_{l,u}\rangle =|\downarrow\downarrow \cdots \chi _{l,u}\cdots \downarrow
\downarrow \rangle$ is prepared.

The magnonic topological pumps proposed in this work could be
efficiently measured in experiments through parallel state preparation
and detection. As illustrated in Fig.~\ref{Fig5}(b), a two-dimensional degenerate
Bose gas of $^{87}$Rb atoms can be prepared in a two-dimensional optical lattice
potential $V(x,y)=V(x)+V(y)$, where $V(x)=V_{\uparrow }(x)+V_{\downarrow }(x)
$ is the state-dependent superlattice potential in the $x$ direction and $%
V(y)=V_{y}\sin ^{2}(2k_{1}y)$ is a state-independent potential in the $y$
direction. When $V_{y}$ is tuned to be vary large, the hopping along the $y$
direction can be ignored. Then $V(x,y)$ creates an array of independent
one-dimensional optical superlattices along the $x$ direction. In the
Mott-insulator regime, this system can be seen as an array of parallel
one-dimensional spin chains described by Eq.~(\ref{DSPC}). One then proposes to simultaneously flip all the spin-down  states located on the left part of the double wells, which form the central column [Fig.~\ref{Fig5}(b)]; this could be realized using a single addressing beam, with a proper linear profile, as was used in a recent experiment on single-magnon states~\cite{Fukuhara2013a}; in this way, extending the entanglement generation procedure described above, each superlattice system could
display a single magnon state,
$|\psi _{l,u}\rangle $. Finally, tuning $\theta $ over one period in
each spin chain, one can realize different topological magnon pumps in parallel, defined in the two-dimensional optical lattice system. In this case, the
magnon density can be efficiently measured by averaging the data extracted
from the magnon density measurements in all spin chains \cite%
{Fukuhara2013a,Fukuhara2013b}.

\section{Conclusion}

In summary, we have proposed and analyzed schemes by which topological insulators and quantized pumps could be realized for magnons, which are excitations that can be addressed and detected in strongly-interacting bosonic systems. Our approach builds on the implementation of the emblematic SSH (resp. RM) models for magnons, using two-component bosonic atoms in state-independent (resp.~dependent) optical superlattices.

We have studied single-magnon quantum dynamics and shown that such nonequilibrium dynamics can be directly exploited to detect topological winding numbers and topological phase transition points. We have also presented an experimentally realistic method to realize quantized topological pumps for magnonic excitations, and we have revealed the importance of the initial (single-magnon) state preparation in this context. We believe that our work will inspire future studies of magnonic topological phases of matter in optical lattice systems.

Importantly, our topological magnonic models are compatible with recent optical-lattice experiments on magnon physics \cite{Fukuhara2013a,Fukuhara2013b}. In this sense, our proposal provides a promising platform for exploring bosonic excitations with topological properties~\cite{Grusdt}. Finally, we note that topological two-magnon bound states have been analyzed in the context of Chern-insulator models~\cite{Qin2017,Qin2018}; it would be interesting to study the properties of such bound states in symmetry-protected one-dimensional models, such as the magnonic SSH model explored in this work. \\

\paragraph*{Acknowledgements}

This work is supported by the National Key R\&D Program of China
(2017YFA0304203); Natural National Science Foundation of China (NSFC)
(11604392, 11674200, 11434007); Changjiang Scholars and Innovative Research
Team in University of Ministry of Education of China (PCSIRT)(IRT\_17R70);
Fund for Shanxi 1331 Project Key Subjects Construction; 111 Project (D18001). NG is supported by the FRS-FNRS (Belgium) and the ERC Starting Grant TopoCold.

\end{document}